\documentclass[%
 reprint,
 superscriptaddress,
 numerical,
 showpacs,
 amsmath,amssymb,
 aps,
 longbibliography,
 pra,
 floatfix
]{revtex4-2}

\usepackage{graphicx}
\usepackage[dvipsnames]{xcolor}

\usepackage[%
  colorlinks   = true, 
  urlcolor     = Blue, 
  linkcolor    = BrickRed, 
  citecolor   = PineGreen 
]{hyperref}

\usepackage[normalem]{ulem}

\usepackage{pifont}

\newcommand{\ket}[1]{| #1 \rangle}
\newcommand{\bra}[1]{\langle#1 |}
\newcommand{\braket}[2]{\langle#1 | #2 \rangle}

\let\Re\relax
\let\Im\relax

\newcommand{\Mtotop}{\hat{\mathcal{M}}} %
\newcommand{\Mtot}{\mathcal{M}} %
\newcommand{\Stot}{\mathcal{S}} %
\newcommand{\Ttot}{\mathcal{T}} %
\newcommand{\Rtot}{\mathcal{R}} %

\DeclareMathOperator{\diag}{diag}
\DeclareMathOperator{\tr}{tr}
\DeclareMathOperator{\sgn}{sgn}
\DeclareMathOperator{\Re}{Re}
\DeclareMathOperator{\Im}{Im}

\def\tcm{T.C.M. Group, Cavendish Laboratory, University of Cambridge, J.J. Thomson Avenue, Cambridge, CB3 0HE, UK}
\def\DAMTP{DAMTP, University of Cambridge, Wilberforce Road, Cambridge, CB3 0WA, UK}

\begin{document}

\title{Surface codes, quantum circuits, and entanglement phases}

\author{Jan Behrends}
\affiliation{\tcm}

\author{Florian Venn}
\affiliation{\DAMTP}

\author{Benjamin B\'eri}
\affiliation{\tcm}
\affiliation{\DAMTP}

\begin{abstract}
Surface codes---leading candidates for quantum error correction (QEC)---and entanglement phases---a key notion for many-body quantum dynamics---have heretofore been unrelated. 
Here, we establish a link between the two. 
We map two-dimensional (2D) surface codes under a class of incoherent  or coherent errors (bit flips or uniaxial rotations) to $(1+1)$D free-fermion quantum circuits via  Ising models.
We show that the error-correcting phase implies a topologically nontrivial area law for the circuit's 1D long-time  state $\ket{\Psi_\infty}$.
Above the error threshold, we find a topologically trivial area law for incoherent errors and logarithmic entanglement in the coherent case.
In establishing our results, we formulate 1D parent Hamiltonians for $\ket{\Psi_\infty}$ via linking Ising models and 2D scattering networks, the latter displaying respective insulating and metallic phases and setting the 1D fermion gap and topology via their localization length and topological invariant.
We expect our results to generalize to a duality between the error-correcting phase of ($d+1$)D topological codes and $d$-dimensional area laws; this can facilitate assessing code performance under various errors.
The approach of combining Ising models, scattering networks, and parent Hamiltonians can be generalized to other fermionic circuits and may be of independent interest.
\end{abstract}

\maketitle

\section{Introduction}

The entanglement of quantum states characterizes many-body phases.
For example, zero modes appear in the entanglement spectrum of topologically ordered~\cite{Li:2008cg,Lauchli:2010jw} and symmetry-protected topological phases~\cite{Pollmann:2010ih,Thomale:2010gi,Turner:2011gp}, including topological insulators and superconductors~\cite{Fidkowski:2010ck}.
The entanglement entropy is a dynamical probe:
for example, starting from local product states, it grows linearly in generic many-body systems, but only logarithmically in many-body localized phases~\cite{Znidaric:2008cr,Bardarson:2012gc}.
Similarly, ground states of gapped local Hamiltonians~\cite{Hastings:2007bu,Eisert:2010hq}, short-range correlated states~\cite{Brandao:2013fz,Brandao:2015cn,Cho_PhysRevX.8.031009}, and almost all many-body localized eigenstates~\cite{Bauer:2013jw} exhibit an area law, i.e., the entanglement entropy grows with a subsystem's area, while generic (random) states follow a volume law~\cite{Page:1993da}.

The entanglement entropy and entanglement spectrum can also characterize purely dynamical phases without an underlying Hamiltonian.
While the long-time evolution of a density matrix with a unitary random circuit will generally yield a volume-law~\cite{Nahum:2017ef,Keyserlingk:2018is}, non-unitary elements change this picture:
When following the quantum trajectory behind a density matrix, i.e., post-selecting measurement outcomes, hybrid circuits that consist of unitary gates and measurements  exhibit a transition between area-law and volume-law phases as a function of measurement rate~\cite{Li:2018bf,Skinner:2019fn,LiPRBet19,Chan:2019ed}.
This transition can also occur in measurement-only dynamics~\cite{Sang:2021fl,Ippoliti:2021jd,Nahum:2020kj,Lang:2020kg}, and similar area-law to logarithmic-law transitions occur for weak measurements~\cite{Cao:2019ef,Chen:2020kd,Alberton:2021fs}.
Due to the post-selection, directly studying these phases experimentally requires a number of runs that is exponential in the system size and circuit depth~\cite{Gullans:2020ed,Ippoliti:2021fu}.
This difficulty may however be overcome via local probes of the entanglement transition~\cite{Gullans:2020ed}, rotating space and time directions in the circuits~\cite{Ippoliti:2021fu,Lu:2021da}, or correlating with classical simulations~\cite{Li:2023fq,JYLee22,Garratt_PhysRevX.13.021026,Garratt2023}.

\begin{figure}
\includegraphics[scale=1]{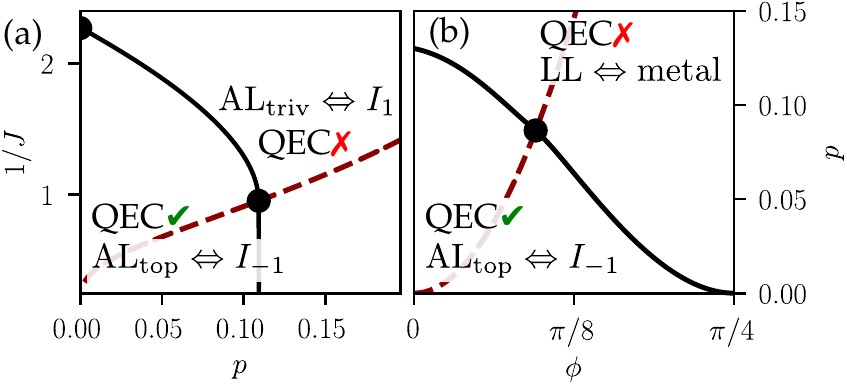}
\caption{Phase diagrams of surface-code QEC, 1D entanglement, and 2D scattering networks. Bit-flip errors give real Ising couplings $J$ [panel (a)]; coherent $\exp[i\phi X]$ errors yield complex couplings $J = -(1/2) \log ( i \tan \phi)$ [panel (b)]. 
The RBIM bond signs, related to $X$ error configurations, are swapped with probability~$p$. 
 The solid lines sketch the phase boundaries, the black dots mark computed phase transition points. 
 The dashed line shows (a) the Nishimori line representing incoherent-error QEC and (b) the $p = \sin^2 \phi$ ``partial Pauli twirl" line for coherent errors . The error-correcting phase (QEC\ding{51}) is dual to a circuit yielding a topologically nontrivial 1D area law with entanglement-spectrum zero modes (AL$_\mathrm{top}$), and to an insulating network with topological invariant $\mathcal{I}=-1$ ($I_{-1}$). Above threshold (QEC\ding{55}) we find (a) a topologically trivial 1D area law (no zero modes, AL$_\mathrm{triv}$) and a $\mathcal{I}=1$ insulator ($I_1$), or (b) a logarithmic entanglement phase (LL) and a metallic network.
}
\label{fig:phase_diagram}
\end{figure}

In this work, we show that entanglement features also usefully characterize quantum error correction (QEC)~\cite{CalShor96,Steane96,Terhal:2015ks}. 
Specifically, we establish a link between entanglement phases in hybrid quantum circuits that we derive from the surface code~\cite{kitaev1997quantum,Kitaev:2003jw,BravyiKitaev_SC,Dennis:2002ds,Fowler12}, and the phases of QEC in the surface code, the latter being a leading candidate for QEC with recent proof-of-principle experiments~\cite{krinner2022realizing,Google_SC}.
Our results are summarized in Fig.~\ref{fig:phase_diagram}.

The link we describe is distinct from recent entanglement--QEC relations via the scrambling of quantum information in hybrid circuits~\cite{hayden2007black,Choi:2020bs,Gullans:2020eg,Li:2021cp,FanPRB21,GullansPRX21,Fidkowski2021howdynamicalquantum,Li:2023fq,Bao:2021hr}.
There, the counter-intuitive robustness of the volume-law phase against a small but non-zero rate of local measurements is explained by this phase supporting emergent QEC code spaces generated by the scrambling dynamics~\cite{hayden2007black,Choi:2020bs,Gullans:2020eg}. 
While this may have potential quantum applications, its uses for fault-tolerant quantum computing have both practical and fundamental limitations~\cite{Hastings2021dynamically,FKNV_review}. 
By focusing on the surface code, the links we describe pertain to codes and error models that are explicitly defined (instead of being emergent), and relate entanglement phases and practically motivated schemes for QEC. 
Our work is also conceptually distinct from topological order emerging in 2+1-dimensional quantum circuits with surface-code ingredients~\cite{Lavasani21,Sang:2021fl}:
Both our concept of interest (i.e., the phases of surface-code QEC) and the quantum circuits this relates to (1+1 instead of 2+1-dimensional circuits as we next note) are different.

Our main result is to map the two-dimensional (2D) surface code under incoherent $X$ errors (i.e., bit-flips) or coherent $\exp[i\phi X]$ errors (with angle $\phi$) to  $(1+1)$D free-fermionic hybrid quantum circuits and to embed the phases of QEC in the entanglement phases of these circuits' long-time 1D states.
This links QEC phases to entanglement phases.
Interestingly, unlike the volume-law--QEC relation in scrambling, we find that the error-correcting phase (QEC phase for short) maps to a 1D area law, namely a topologically nontrivial 1D phase.

The overture to establishing these links is a mapping from the 2D surface code to 2D random-bond Ising models (RBIMs)~\cite{Dennis:2002ds,Venn2022}. 
This opens a direct route to $(1+1)$D dynamics upon viewing the RBIM transfer matrix~\cite{Schultz:1964fv,Cho:1997bs,Merz:2002gj} as a quantum circuit.
While for bit flips, yielding real Ising couplings~\cite{Dennis:2002ds}, this is the familiar $(d+1)$D classical to $d$-dimensional quantum duality, additional considerations are needed for the coherent case where Ising couplings are complex~\cite{Dennis:2002ds,Venn2022}. 
This is provided by a further mapping between 2D Ising models and 2D scattering networks~\cite{Cho:1997bs,Merz:2002gj,Venn2022}. 

The entanglement phases of QEC, and the broader entanglement (and Ising) phases they are embedded in, are sketched in Fig.~\ref{fig:phase_diagram}:
For bit-flips [real Ising couplings, panel (a)], we find area-law phases both below and above the QEC threshold. These phases correspond to an insulating network (see also Refs.~\onlinecite{Dennis:2002ds,Merz:2002gj,Venn2022,Jian:2022jg}).
The nontrivial long-time-state topology below threshold is signaled by a zero mode in the entanglement spectrum and by the (interrelated) topological invariants for the 1D state and the 2D network~\cite{Merz:2002gj,Venn2022}.
For coherent errors [complex Ising couplings, panel (b)], the QEC phase corresponds to the same entanglement phase (and network-model phase, cf.\ Ref.~\onlinecite{Venn2022}) as the incoherent QEC phase. Above threshold, however, we find a phase with entanglement entropy increasing logarithmically with system size. Here the network is metallic (see also Ref.~\onlinecite{Venn2022}).

While these results build on two existing links, from surface code QEC to Ising and network models on the one hand~\cite{Dennis:2002ds,Merz:2002gj,Venn2022} and between network models and entanglement phases on the other~\cite{Jian:2022jg}, they establish a  conceptually novel link between surface code QEC and entanglement phases in transfer matrix space, a connection we expect to exemplify a broader correspondence with intriguing implications~\cite{Bravyi:2014ja}.
Firstly, our construction naturally generalizes to dualities between $(d+1)$D codes with a local structure (such as topological codes~\cite{BombinTQC,Terhal:2015ks}) and $d$-dimensional entanglement phases in $(d+1)$D quantum circuits.
Secondly, by having found it to emerge for qualitatively different error types (bit flips and coherent errors), we expect a general correspondence between the QEC phase and transfer matrix  area laws.
This opens the door to using the area laws' classical simulability to chart the QEC phase for various codes and errors, including settings with coherent errors where a free-fermion description is unavailable, thus tackling a key challenge in QEC~\cite{Kueng16,Wallman16,Debroy18,Bravyi:2018ea,iverson2020coherence,Hashim21}.
Furthermore, by mapping error-corrupted codes to the entanglement structure arising from the long-time (i.e., infrared) dynamics of a system one dimension lower, our results anticipate a deep connection to the characterization of error-corrupted topologically ordered states via boundary phases and their transitions~\cite{Bao2023,Fan2023,Lu:2023ep,Zou:2023he}.

On a technical level, our use of Ising and network models together allows us to analytically establish the correspondence between Anderson insulating (i.e., disordered) networks and free-fermion area-laws, which we further characterize via quasi-local 1D parent Hamiltonians. These advances, extending clean-system analytic results and disorder numerics~\cite{Jian:2022jg}, may be of independent interest for free-fermion hybrid quantum circuits.
%


\section{Ising models for random bit flips and for coherent errors}

The link between surface-code QEC and hybrid quantum circuits will be through a generalized 2D RBIM~\cite{Schultz:1964fv,Cho:1997bs,Merz:2002gj,Venn2022} on the square lattice, with Hamiltonian
\begin{align}
 H = -\sum_{\langle v ,v \rangle} J \eta_{vv'} \sigma_v \sigma_{v'}
 \label{eq:ising_hamiltonian}
\end{align}
and partition function $Z = \sum_{ \{ \sigma_v \} } \exp ( -H)$ (the inverse temperature is absorbed into the couplings).  The Ising spins $\sigma_v = \pm 1$  have nearest-neighbor coupling constant $J$ with random signs $\eta_{vv'}$; the latter are drawn from an uncorrelated random distribution where $\eta_{vv'} =1$ with probability $1-p$ and $\eta_{vv'}=-1$ with probability~$p$.
We consider two choices of $J$: either purely real or complex $J = -(1/2) \log (i \tan \phi)$ with $\phi \in [0,\pi/4]$.

\vspace*{1em}

\subsection{Surface code basics}

As we explain below, both choices for the couplings in Eq.~\eqref{eq:ising_hamiltonian} originate in surface-code QEC~\cite{Dennis:2002ds,Venn2022}, cf.\ Fig.~\ref{fig:setup}. 
We consider the 2D toric code~\cite{Kitaev:2003jw,Kitaev:1997kq} on the square lattice.
This is a topological stabilizer code~\cite{Terhal:2015ks} with qubits on the lattice links and with two types of stabilizers that (in the bulk) each act on four neighboring qubits:
$X$-stabilizers $S_v^X = \prod_{i \in v} X_i$ are assigned to vertices $v$ and $Z$-stabilizers $S_w^Z = \prod_{i \in w} Z_i$ to plaquettes $w$, where $X_i$ and $Z_i$ are Pauli operators~\cite{Kitaev:2003jw}.
The states $\ket{\psi}$ that for all $v$ and $w$ satisfy $S_v^X \ket{\psi}=\ket{\psi}$ and $S_w^Z \ket{\psi}=\ket{\psi}$ constitute the logical subspace.
The number of states that satisfy these conditions depends on the boundary conditions~\cite{Kitaev:2003jw}. 
For concreteness, we focus on a cylinder geometry with boundary conditions yielding a two-dimensional computational space, i.e., one logical qubit.
(Our considerations, however, are more general, cf.\ Ref.~\onlinecite{Venn2022} for details on a planar geometry.)
In particular we use ``smooth boundaries"~\cite{Dennis:2002ds} so that one of the logical operators is $\bar{X} = \prod_{i \in \gamma} X_i$ with the product being over qubits $i$ on the shortest sequence $\gamma$ of vertical links (in terms of Fig.~\ref{fig:setup}) along the length of the cylinder, while the conjugate logical operator $\bar{Z} = \prod_{i \in \gamma'} Z_i$ is the product of $Z_i$ along its circumference (with the shortest path $\gamma'$ of vertical links). 
Equivalent logical operators arise from these upon stabilizer multiplication.

The toric code can correct $X$-errors and $Z$-errors independently; the considerations for the two are analogous. 
In what follows, we focus on $X$-errors, considering incoherent bit flips, i.e., $X_i$ being applied with probability $p$, or coherent errors from the application of $\exp[i\phi X_i]$ on each qubit. 
(The latter arise from unwanted gate rotations---ubiquitous in quantum devices.)
A string of $X_i$ being applied, whether from bit-flips or as a contribution from $\prod_i\exp[i\phi X_i]$,
can be detected by syndrome measurements: $S_w^Z =-1$ mark the end points of applied $X_i$-strings. 
The set of $S_w^Z$ eigenvalues is called the syndrome $s$.
Given syndrome $s$, applying an $X_i$ string $C_s$ with the same end points returns the state to the computational space~\cite{Dennis:2002ds}.
While the end points are fixed, the strings $C_s$ themselves can vary: Applying $S_v^X$ to $C_s = \prod_{i \in s} X_i$ adds or removes loops of $X_i$ operators, and thus changes the strings contained in $C_s$ but not their end points. 
Furthermore, by $S_v^XC_s=C_sS_v^X$ and $S_v^X\ket{\psi}=\ket{\psi}$, this leaves $C_s\ket{\psi}$ invariant.
Applying $\bar{X}$, however, also leaves the end points invariant, but $\bar{X}C_s\ket{\psi}\neq C_s\ket{\psi}$.
Hence there are two inequivalent classes (homology classes~\cite{Dennis:2002ds}) of error: those equivalent to  $C_s$ and those to $\bar{X}C_s$. 
In QEC, given syndrome $s$, a decoder must decide which homology class the error is in and hence whether to apply $C_s$ or $C_s \bar{X}$ to return the state back to the logical subspace.
We denote both cases by $C_s \bar{X}^q$ with $q=0,1$.

\subsection{Ising mappings}
\label{sec:Ising_mappings}

We now relate surface code QEC to Eq.~\eqref{eq:ising_hamiltonian}, starting with random bit flips (i.e., incoherent errors~\cite{Nielsen:2010ga,Haake:2010hc}). 
For this case, we follow Ref.~\onlinecite{Dennis:2002ds}.
On each qubit $j$, a bit flip $X_j$ occurs with probability $p$; the qubit stays intact with probability $1-p$. 
Thus the probability of an $X$-string $C_s \bar{X}^q$ occurring is the product 
\begin{equation}
P_{C_s,q}=\prod_j \sqrt{p(1-p)} e^{\eta^{(C_s,q)}_j J},\quad e^{J}=\sqrt{\frac{1-p}{p}}, 
\label{eq:Pstringq}
\end{equation}
over all qubits where $\eta^{(C_s,q)}_j=-1$ if $X_j$ occurs (i.e., contained in $C_s \bar{X}^q$) and $\eta^{(C_s,q)}_j = 1$ otherwise. 
Henceforth we suppress the superscripts in $\eta^{(C_s,q)}_j$.

As we noted above, a syndrome $s$ does not determine a unique string $C_s$, but only its end points.
To obtain the probability $P_{s,q}$ that syndrome $s$ occurs, and does so via an error in the homology class $q$ of $C_s \bar{X}^q$, we thus need to sum over $P_{C'_s,q}$ for all other strings $C'_{s} \bar{X}^q$ with the same $s$ and $q$. 
Fixing a reference string $C_s \bar{X}^q$ and multiplying it by $X$-stabilizers generates another such string, $C^\prime_{s} \bar{X}^q = \prod_v (S_v^X)^{n_v} C_s \bar{X}^q$, where $n_v \in \{0, 1\}$. 
The set of all $\{n_v\}$ configurations generates all such homologically equivalent strings, i.e., all strings given $s$ and $q$.

\begin{figure}
\includegraphics[scale=1]{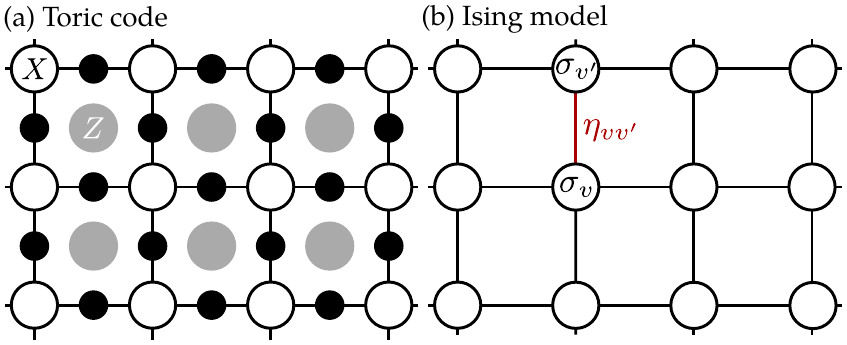}
\caption{(a) A bulk patch of the toric code, with physical qubits marked by black dots, while $X$- and $Z$-stabilizers by white and gray disks, respectively. (b) The code is mapped to a RBIM with real couplings for incoherent and complex couplings for coherent errors; $X$-stabilizers map to Ising spins $\sigma_v$. The nearest-neighbor couplings have sign $\eta_{vv'}$.}
\label{fig:setup}
\end{figure}

On each vertex, we now introduce Ising spins $\sigma_v = (-1)^{n_v} = \pm 1$ (valued $-1$ when $S_v^X$ is contained in the stabilizer product and $1$ otherwise), cf.\ Fig.~\ref{fig:setup}.
Each qubit has two neighboring $X$-stabilizers. 
(This is evident in the bulk from Fig.~\ref{fig:setup}; we also use cylinder termination with this property~\cite{Venn2022}.)
Each qubit thus corresponds to a bond between nearest neighbor $\sigma_v$.
When exactly one $X$-stabilizer neighboring qubit $j$ is contained in the stabilizer product, we must swap $p \leftrightarrow 1-p$  in the corresponding factor in $P_{C'_s,q}$; in order words, the exponent $e^{\eta_j J} \to e^{-\eta_j J}$.
We can express this swap via the $\sigma_v$: For each qubit, i.e., Ising bond, we include the product of the two neighboring $\sigma_v$ and thus write the product over all qubits as one over all Ising bonds
\begin{equation}
P_{C'_s,q}=\prod_{\langle v,v'\rangle} \sqrt{p(1-p)} e^{\eta_{vv'} J \sigma_v \sigma_{v'}}.
\end{equation}
Here, we also relabeled $\eta_{j} \to \eta_{vv'}$, using again that each qubit is located at bonds between nearest-neighbor $\sigma_v$.

Summing over all possible strings with a given $s$ and $q$, or equivalently over all Ising spin configurations $\{ \sigma_v \}$, we obtain $P_{s,q} = (\prod_j \sqrt{p(1-p)}) Z_{s,q}$, where $Z_{s,q} = \sum_{\{ \sigma_v \} } \exp (-H_{s,q})$ is the RBIM partition function. 
Here, $H_{s,q}$ defined in Eq.~\eqref{eq:ising_hamiltonian} with $J=(1/2) \log ((1-p)/p)$; 
the subscripts $s,q$ denote the reference string $C_{s}\bar{X}^{q}$ that sets the configuration $\{ \eta_{vv'} \}$ for $H_{s,q}$ [implicit in Eq.~\eqref{eq:ising_hamiltonian}].
Since random bit flips occur with probability $p$, the signs $\eta_{vv'} = -1$ with probability~$p$ and $\eta_{vv'} = 1$ with probability $1-p$. 
This choice of $p$ and $J$ defines the Nishimori line~\cite{Dennis:2002ds,Nishimori:1981dw}.
Our exploration of entanglement phases includes both this line, but we are also interested in the phase diagram in the broader $p-J$ space [cf.\ Fig.~\ref{fig:phase_diagram}(a)].

We next review, following Ref.~\onlinecite{Venn2022}, the Ising mapping for coherent errors of the form $U= \prod_i U_i$ with $U_i = \exp ( i \phi X_i )$~\cite{Shor:1996jj}.
The probability $P_{s,q}$ now arises from an overlap, $P_{s,q} = | \bra{\psi} C_s \bar{X}^q U \ket{\psi} |^2$. 
Here, we take $\ket{\psi}$ to be the $+1$ eigenstate of $\bar{Z}$ so that $\ket{\psi}$ and $\bar{X}\ket{\psi}$ are orthogonal and hence $P_{s,q}$ are probabilities. 
(This $P_{s,q}$ is also related a QEC fidelity under a suitable Bloch-sphere average~\cite{Venn:2020ge,Venn2022}.)
The amplitude $\bra{\psi} C_s \bar{X}^q U \ket{\psi}$ can be evaluated similarly to how $P_{s,q}$ was in the incoherent case, but now instead of a sum over the probabilities $P_{C'_s,q}$ of various $X$-strings, the expansion of $U$ involves their  coherent sum. 
To get the amplitude, we must thus replace $p\to i\sin(\phi)$ and $1-p\to \cos(\phi)$ in our previous derivation.
As a result, $P_{s,q} = (\prod_j |\sin \phi \cos \phi|) |Z_{s,q}|^2$ with  $J= -(1/2) \log (i \tan \phi)$. 
As before, $Z_{s,q} = \sum_{\{ \sigma_v \} } \exp (-H_{s,q})$ with the $H$ of Eq.~\eqref{eq:ising_hamiltonian}.

For the coherent-error QEC problem, now $\phi$ sets the syndrome distribution and hence $\eta_{vv'}$, in a coherent generalization of the Nishimori line~\cite{Venn2022}. 
Sampling $\eta_{vv'}$ according to this is more difficult than for bit flips: 
instead of sampling independently for each qubit (i.e., Ising bond), one must now sample bonds in certain sequence~\cite{Bravyi:2018ea,Venn:2020ge} to sample from $P_s = P_{s,0} + P_{s,1}$~\cite{Bravyi:2018ea,Venn:2020ge}.
While this is needed for quantitative accuracy (e.g., for the error threshold or for critical properties), here we use a simplified model where we draw $\eta_{vv'}$ from an uncorrelated distribution with $\eta_{vv'} =-1$ occurring with probability $p$ and $\eta_{vv'}=1$ with probability $1-p$.
This model thus has a $\phi-p$ phase diagram [Fig.~\ref{fig:phase_diagram}(b)].

Taking $p = \sin^2 \phi$ [shown dashed in Fig.~\ref{fig:phase_diagram}(b)] in this $\phi-p$ space  mimics the QEC problem in a manner reminiscent of the Pauli twirl approximation~\cite{Emerson:2007ej,Silva:2008ij} which replaces each $U_i$ by a bit flip occurring with probability $p = \sin^2 \phi$. 
Pauli twirling would however make this replacement from the outset, yielding the incoherent RBIM at this $p$, in contrast to using $p = \sin^2 \phi$ with the complex RBIM. 
The latter goes qualitatively beyond Pauli twirling: the partition function (and hence the quantum circuit below) accounts for the coherent sum over $X$-string amplitudes---the key feature distinguishing coherent from incoherent errors.
For this reason, we call $p = \sin^2 \phi$ in the complex RBIM a ``partial Pauli twirl". 

Along the $p = \sin^2 \phi$ partial Pauli twirl line we expect the qualitative structure of the QEC phase diagram to be the same for the full coherent error model and our simplified model; we shall further substantiate this expectation in Sec.~\ref{sec:network_model} using the scattering network description.

\section{Quantum circuit}

To relate the Ising models to quantum circuits, we express the partition function using the transfer matrix~\cite{Schultz:1964fv,Merz:2002gj}.
Following standard steps~\cite{Cho:1997bs,Read:2000hu,Merz:2002gj,Schultz:1964fv,sachdev_2011,fradkin2013field} agnostic to whether couplings are real or complex, the partition function for a system on a cylinder of length $L$ and circumference $M$ is $Z=\bra{\alpha_L}\mathcal{M}\ket{\alpha_0}$ where $\ket{\alpha_{r}}$ encodes boundary conditions at the $x=0,L$ ends of the cylinder and $\Mtotop$ is the transfer matrix
\begin{equation}
 \Mtotop=V_L H_{L-1} \dots      H_2 V_2 H_1 V_1,
 \label{eq:Mtotop}
\end{equation}
where the hat distinguishes this many-body operator from its single-particle counterpart $\Mtot$ in Sec.~\ref{sec:network_model}.
The two kinds of transfer matrix layers are
\begin{align}
 H_n &= \prod_{i=1}^M A_{n,i} \exp \left( \sum_{i=1}^{M} \tilde{\kappa}_{n,i} X_i \right) ,~A_{n,i} = \sqrt{ \frac{2}{\sinh (2 \tilde{\kappa}_{n,i})} } \nonumber \\
 V_n &= \exp \left( -\sum_{i=1}^{M} \kappa_{n,i}  Z_i Z_{i+1} \right), \label{eq:ising_transfer} 
\end{align}
with  $\kappa_{n,i} = J \eta_{n,i}^{(v)}$ and $\tilde{\kappa}_{n,i} = -(1/2) \log [\tanh ( J \eta_{n,i}^{(h)} )]$ where the labels $h$ and $v$ distinguish horizontal ($\eta_{n,i}^{(h)}$) and vertical Ising bonds ($\eta_{n,i}^{(v)}$).
The Pauli $X_i$ and $Z_i$ in Eq.~\eqref{eq:ising_transfer} act in an $M$-site 1D transfer matrix space.
The $V_i$ and $H_i$ involve complementary terms from the transverse field Ising model. 
The layers thus commute with the $\mathbb{Z}_2$ symmetry $P=\prod_j X_j$ shared with this model.

\begin{figure}
\includegraphics[scale=1]{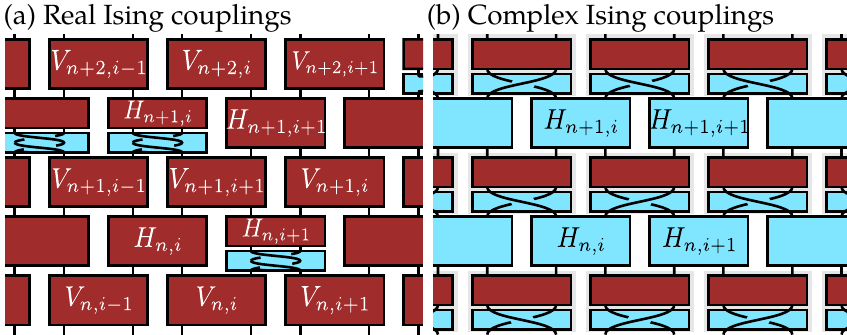}
\caption{Fermionic quantum circuits acting on Majorana fermion lines. The ``time" direction, along the cylinder, is upwards. (a) Circuit for real Ising couplings $J$ (from incoherent bit flips). If $\eta_{n,i}^{(h)}<0$, the $H_{n,i}$ gates involve unitary double-braids (blue) and imaginary time evolution (dark red). (b) For complex Ising couplings $J = -(1/2) \log (i \tan \phi)$, from coherent errors, the gates $H_{n,i}$ are always unitary (blue), while the $V_{n,i}$ consist of unitary braids (blue) and imaginary time evolution (dark red).
}
\label{fig:circuit}
\end{figure}

Since the individual terms in the exponentials in Eqs.~\eqref{eq:ising_transfer} mutually commute, we write $H_n = \prod_{i} A_{n,i} H_{n,i}$ with gates $H_{n,i} = \exp \left( \tilde{\kappa}_{n,i} X_i \right)$ and  $V_n = \prod_{i}^M V_{n,i}$ with gates $V_{n,i} = \exp \left( -\kappa_{n,i}  Z_i Z_{i+1} \right)$.
The transfer matrix thus consists of a successive application of layers of one-body and two-body gates: it is a quantum circuit. 
The gates are not unitary, but depending on whether $J$ is real or complex, they can yield both real and imaginary time evolution (see below and Fig.~\ref{fig:circuit}).
We define the entanglement phases of $\Mtotop$ as of the long-time state $\ket{\Psi_\infty}$ obtained by time-evolution with $\Mtotop$, starting a generic definite-parity initial state $\ket{\Psi_0}$ (not necessarily $\ket{\alpha_0}$). 

It will be beneficial to explore this in a fermionic setting. 
This will allow us to show that $\Mtotop$ is (essentially) a free-fermion circuit and that $\ket{\Psi_0}$ can be taken as a fermionic Gaussian state, without loss of generality. 
(Gaussian states are ground states or thermal states of free-fermion Hamiltonians~\cite{Bravyi:2005jh,Fidkowski:2010ck}.)
To construct a fermionic quantum circuit from $\Mtotop$, we switch to a Majorana basis~\cite{Merz:2002gj} via a Jordan-Wigner transformation~\footnote{We use the convention $\gamma_{2i-1} = \prod_{k<i} X_k Z_i$, $\gamma_{2i} = -\prod_{k<i} X_k Y_i$.}. 
The Majorana fermions $\gamma_i^\dagger = \gamma_i$ ($i=1\ldots 2M$) satisfy the canonical anticommutation relations $\{ \gamma_i,\gamma_j \} = 2 \delta_{ij}$~\cite{Kitaev:2001gb}, and allow one to express the parity as $P = (-i)^M \gamma_1 \gamma_2 \dots \gamma_{2M}$. 
The gates are now $H_{n,i} = \exp \left( -i \tilde{\kappa}_{n,i} \gamma_{2i-1} \gamma_{2i} \right)$, $V_{n,i<M} = \exp \left( -i \kappa_{n,i}  \gamma_{2i} \gamma_{2i+1} \right)$, and $V_{n,M} = \exp \left( i P \kappa_{n,M} \gamma_{2M}\gamma_{1} \right)$.

The appearance of $P$ in $V_{n,M}$ is due to the nonlocality of the Jordan-Wigner transformation; it arises from describing a bosonic, i.e., qubit-based, quantum circuit with fermions.
As swapping $P$ swaps the sign of a fermionic hopping around the cylinder, changing the fermion parity changes between periodic and antiperiodic boundary conditions (pbc and apbc, respectively) on fermions. 
[Note that an $\eta_{n,i}^{(v)}$ string corresponding to $\bar{X}$ achieves the same, so changing $q$ also changes between pbc and apbc.]
While retaining $P$ in $V_{n,M}$, and hence the intertwined parity and fermion boundary conditions, is important for establishing surface code QEC features from the circuit~\cite{Venn2022}, 
it is less crucial for establishing the circuit's entanglement phases, provided we consider both pbc and apbc and both parities for fermions (see Secs.~\ref{sec:FSGS} and \ref{sec:GGSBC}). 
In this way, we can also use $V_{n,M} = \exp \left( -i  \kappa_{n,M} \gamma_{2M}\gamma_{1} \right)$. 
Henceforth we call circuits with this $V_{n,M}$ ``purely fermionic", to distinguish from the fermionized transfer matrix $\Mtotop$ (henceforth called ``bosonic'').
 
These quantum circuits are different from previous fermionic mappings of the toric code~\cite{Wen:2003eg,Bravyi:2018ea}: these involve Abrikosov pseudo-fermions~\cite{Abrikosov:1965go} that have a parity constraint.
While the pseudo-fermion mapping can be used to sample from the syndrome probabilities in the coherent case~\cite{Bravyi:2018ea}, the circuit that arises from a statistical-mechanics mapping computes the probabilities for the different homology classes, and incorporates both incoherent and coherent errors on a unified footing.

We now discuss how the circuit combines real and imaginary time evolution, cf.\ Fig.~\ref{fig:circuit}. 
Real couplings $J$ (from incoherent bit flips) correspond to purely imaginary time evolution up to double braiding of Majoranas:
The $V_{n,i}$ are matrix exponentials of Hermitian operators, but for the $H_{n,i}$ this is true only when $\eta_{n,i}^{(h)}=1$. 
When $\eta_{n,i}^{(h)}=-1$, $H_{n,i} = \gamma_{2i-1}\gamma_{2i} \exp (-i \sum_i \Re (\tilde{\kappa}_{n,i} ) \gamma_{2i-1}\gamma_{2i} )$, where $\gamma_{2i-1}\gamma_{2i}$ is the unitary double braiding of Majoranas, cf.\ Fig.~\ref{fig:circuit}(a)~\cite{Aasen_PhysRevX.6.031016,Martin_PRXQuantum.1.020324,BehrendsSYKcirc}.
Complex couplings $J = -(1/2) \log (i \tan \phi)$ (from coherent errors) correspond to a mixed real- and imaginary time evolution: Here, the $H_{n,i}$ are unitary operators since $\tilde{\kappa}_{n,i} = i (\phi - (1-\eta_{n,i}^{(h)}) \pi/4)$ is purely imaginary.
The operator $V_{n,i}$ can be decomposed into unitary braiding $(1-\eta_{n,i}^{(v)} \gamma_{2i} \gamma_{2i+1})/\sqrt{2}$ and imaginary time evolution $\exp ( - i \Re \kappa_{n,i}  \gamma_{2i} \gamma_{2i+1})$, as illustrated in Fig.~\ref{fig:circuit}(b).
The surface code, in particular with coherent errors, thus provides a concrete physical motivation for fermionic quantum circuits alternating real and imaginary time evolution, studied in relation to emergent conformal symmetries~\cite{Chen:2020kd} and classifications of fermionic quantum circuits and tensor networks~\cite{Jian:2022jg}.

\subsection{Final states as 1D ground states}
\label{sec:FSGS}

We now further specify the settings for defining the entanglement phases of $\Mtotop$. 
We consider the properties of a long-time state $\ket{\Psi_\infty}$.  
That is, we consider the large $L$ limit of the evolution
\begin{equation}
 \ket{\Psi_L} = H_L V_L \dots H_2 V_2 H_1 V_1 \ket{\Psi_0}, 
\end{equation}
where normalizing $ \ket{\Psi_L}$ (as required by the evolution not being unitary) is left implicit. 

As $\Mtotop$ is nonunitary, a useful view on its features can be obtained from its singular value decomposition. We write~\cite{Merz:2002gj,Venn2022}
\begin{equation}\label{eq:SVD}
\Mtotop=\sum_n e^{-E_nL/2}\ket{\varphi_n}\!\bra{\widetilde{\varphi}_n}, 
\end{equation}
where the left singular vectors $\ket{\varphi_n}\!$ are the eigenvectors of $\Mtotop \Mtotop^\dagger$ and the right singular vectors $\ket{\widetilde{\varphi}_n}$ are the eigenvectors of $\Mtotop^\dagger\Mtotop$.
The energies $E_n$ can be interpreted as those of a 1D Hamiltonian $\mathcal{H}$, defined by $\Mtotop \Mtotop^\dagger=e^{-L\mathcal{H}}$, which has $\ket{\varphi_n}\!$ as its eigenvectors. 

We next define the large $L$ limit more carefully: Considering that $\Mtotop$ is parity conserving, and denoting by $\delta\varepsilon$ the gap between the lowest and second-to-lowest energies of $\mathcal{H}$ eigenstates with the same parity as that of $\ket{\Psi_0}$, we define the large $L$ limit by $L\delta\varepsilon\gg1$. (The energy levels of $\mathcal{H}$ become increasingly non-random upon increasing $L$, cf.\ Sec.~\ref{sec:2Dnet1DH}.)
For $\ket{\Psi_\infty}$, this implies
\begin{equation}
\ket{\Psi_\infty}= e^{-E_\text{min} L/2}\ket{\varphi_\text{min}}\!\braket{\widetilde{\varphi}_\text{min}}{\Psi_0},
\label{eq:long-time-limit}
\end{equation}
hence $\ket{\Psi_\infty}\propto \ket{\varphi_\text{min}}$,  the lowest-energy state (with energy $E_\text{min}$) of $\mathcal{H}$ with the same parity as that of $\ket{\Psi_0}$. 
This is the ground state of $\mathcal{H}$ (or a ground state if there is a degenerate ground space) only if a ground state exists with this parity.
This distinction is important when $\mathcal{H}$ is gapped. 
In this case, we consider states $\ket{\Psi_0}$ with each parity and, depending on whether we deal with the fermionized bosonic $\Mtotop$ or its purely fermionic version, we also consider both pbc and apbc, i.e., $q=0,1$ (see Sec.~\ref{sec:GGSBC}). 
In this way, when $\mathcal{H}$ is gapped, we can take $\ket{\Psi_\infty}$ to be a ground state. 
(This is easily identifiable by the fast exponential convergence due to the gap.) 
That is, when $\mathcal{H}$ is gapped, by the entanglement phases of $\Mtotop$  we mean those of this ground-state-converged $\ket{\Psi_\infty}$. 
(When $\mathcal{H}$ is gapless we do not need such qualification because $\ket{\Psi_\infty}$ is similar for either parity.)

\subsection{Gaps, ground states, boundary conditions}
\label{sec:GGSBC}

For the characterization of $\ket{\Psi_\infty}$, a further key feature is that the gates $V_{n,i}$ and $H_{n,i}$ are quadratic, and hence $\mathcal{H}$ is a 1D free fermion Hamiltonian. 
(This holds as is for the purely fermionic version of $\Mtotop$; for the bosonic $\Mtotop$ it holds for each parity.)
This implies that $\ket{\Psi_\infty}$ is a free-fermion state for any definite-parity initial state $\ket{\Psi_0}$. 
Viewing $\ket{\Psi_\infty}$ as a free-fermion ground state is particularly useful in establishing its topological and entanglement features.

To establish a topological characterization, we will use that gapped free-fermion Hamiltonians in 1D are distinguished by the response of their ground-state fermion parity to a change between pbc and apbc. 
Specifically, in a topologically nontrivial system, the respective ground-state parities satisfy $P_\text{GS}^\text{pbc}=-P_\text{GS}^\text{apbc}$~\cite{Kitaev:2001gb}. 
In a topologically trivial system we have $P_\text{GS}^\text{pbc}=P_\text{GS}^\text{apbc}$. 
This allows one to define a topological invariant $\mathcal{I}=P_\text{GS}^\text{pbc}P_\text{GS}^\text{apbc}$ with $\mathcal{I}=-1$ in a topological phase~\cite{Kitaev:2001gb}.

The topological aspects and boundary conditions are thus intertwined. 
In particular, while the ground state is always unique for the purely fermionic gapped $\mathcal{H}$, subtleties arise in the bosonic problem when $\mathcal{I}=-1$. 
This is because in this problem, changing $P$ changes the fermion boundary conditions and for $\mathcal{I}=-1$, changing these boundary conditions changes $P_\text{GS}$, where $P_\text{GS}$ is the ground-state parity of the purely fermionic $\mathcal{H}$. 
Hence, either $P=P_\text{GS}$ for both $P=\pm1$ or for neither. 
That is, when the purely fermionic $\mathcal{H}$ is gapped and has $\mathcal{I}=-1$, the state $\ket{\varphi_\text{min}}$ in Eq.~\eqref{eq:long-time-limit} for the bosonic problem is either a ground state of this purely fermionic $\mathcal{H}$ for both $P=\pm1$, or it is its lowest excited state for both $P=\pm1$. 
The two-fold ground space degeneracy in the former case, of course, just corresponds to spontaneous symmetry breaking in the spin-chain, generalizing that in the transverse-field Ising chain.

In this $\mathcal{I}=-1$ case, one can switch between $\ket{\varphi_\text{min}}$ being a ground or excited fermionic state by switching $P_\text{GS}$ (without changing $P$). 
This is achieved by changing boundary conditions via changing $q$, i.e., changing $\eta_{n,i}^{(v)}$ along the cylinder, corresponding to the application of $\bar{X}$. 
(For $\mathcal{I}=1$, both $q$-values work because $P_\text{GS}^\text{pbc}=P_\text{GS}^\text{apbc}$.)
The above considerations highlight that, depending on whether we use a purely fermionic or the bosonic form of the circuit $\Mtotop$, exploring the entanglement phases requires considering both parities and boundary conditions. 
(In practice, choices exist that work for most disorder realizations. For example, $q=0$ for the bosonic $\Mtotop$ works because $\mathcal{I}=-1$, as we will show, arises for small $p$ where long $\eta_{n,i}^{(v)}=-1$ chains, effecting a spurious boundary condition change to be undone by $q=1$, have probability exponentially suppressed in $L$.)

\subsection{Characterizing free-fermion entanglement}
The quantum circuit having quadratic gates also enables both the single-particle characterization and the efficient numerical evaluation of entanglement properties.
Using that $\ket{\Psi_\infty}$ is the same Gaussian state regardless of the details of the definite-parity initial state $\ket{\Psi_0}$, we can choose $\ket{\Psi_0}$ to be Gaussian as well. 
We can then use that any Gaussian state evolved by $H_n$ and $V_n$ remains Gaussian~\cite{Bravyi:2005jh}, with the same parity as that of $\ket{\Psi_0}$. 
This implies that all many-body quantities can be computed using fermionic linear optics~\cite{Bravyi:2005jh}.
The central object of this approach is the correlation matrix
\begin{equation}
 C_{jk} = \frac{i}{2} \tr \left( \rho [\gamma_j , \gamma_k ] \right)
\end{equation}
from which all higher correlators follow~\cite{Bravyi:2005jh}.
Following \citeauthor{Bravyi:2005jh}, we evolve the matrix $C_{ij}^{(n)} \to C_{ij}^{(n+1)}$ directly~\cite{Bravyi:2005jh} (with $n$ denoting the time step) instead of the (exponentially large) density matrix $\rho^{(n)} \to \rho^{(n+1)} = T_n \rho^{(n)} T_n^\dagger / \tr[ T_n \rho^{(n)} T_n^\dagger ]$ with $T_{2n} = H_n$ and $T_{2n-1} = V_n$---cf.\ Appendix~\ref{sec:appendix_evolution} for more details.

To calculate the entanglement entropy and entanglement spectrum for a subsystem $A$, the correlation matrix $\bar{C}_A^{(n)}$ of the corresponding reduced density matrix $\bar{\rho}_A^{(n)}$ (also a Gaussian state~\cite{Bravyi:2005jh,Fidkowski:2010ck}) can be obtained from $C^{(n)}$ by keeping only those indices contained in $A$.
Since $\bar{C}_A^{(n)}$ is real and antisymmetric, it can be block-diagonalized via a Youla decomposition~\cite{YoulaCJM1961} $\bar{C}^{(n)}_A = Q \Sigma Q^T$ where $\Sigma = \diag ( \{ i \lambda_r Y \} )$ and $Y$ is the 2\textsuperscript{nd} Pauli matrix.
The set of $\lambda_r$ is the single-particle entanglement spectrum~\cite{Fidkowski:2010ck}; the matrix $i\bar{C}_A^{(n)}$ is the  single-particle ``entanglement Hamiltonian". 
It determines the entanglement entropy as
\begin{equation}
 S_A = - \sum_r \frac{1-\lambda_r}{2} \log \frac{1-\lambda_r}{2} -\sum_r \frac{1+\lambda_r}{2} \log \frac{1+\lambda_r}{2}.
\end{equation}
We will from now on consider the entanglement spectrum and entropy only for bipartitions of the system into two halves of size $M/2$, and denote the entanglement entropy by $S_{M/2}$.
As we shall see in the Sec.~\ref{sec:results}, the entanglement spectrum and the entanglement entropy are key characteristics of the circuit $\Mtotop$ and hence also characterize surface-code QEC. 

\section{Network model}
\label{sec:network_model}

\begin{figure}
\includegraphics[scale=1]{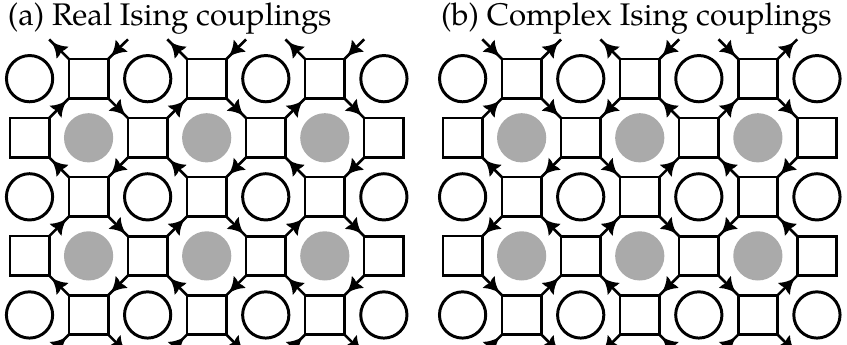}
\caption{Network model for the (a) real and (b) complex Ising couplings from incoherent and coherent errors, respectively. At each Ising bond in Fig.~\ref{fig:setup}(b) we now have junction matrices $H_{n,i}$ and $V_{n,i}$. They scatter directed Majorana modes residing on the networks' links. Translating $H_{n,i}$ and $V_{n,i}$ into junction scattering matrices requires different link direction layouts for real and complex Ising couplings: counter-propagating pairs of links for real couplings~\cite{Merz:2002gj} and co-propagating pairs for complex Ising couplings~\cite{Venn2022}.
}
\label{fig:network_model}
\end{figure}

We now turn to the network model (cf.\ Fig.~\ref{fig:network_model}).
For a many-body operator $T= \exp (i \sum_{ij} \gamma_i q_{ij} \gamma_j) $, single Majorana operators transform as~\cite{Kitaev:2001gb}
\begin{equation}
T \gamma_i T^{-1} = \sum_j t_{ji} \gamma_j,\quad t = \exp (4 i q).
\label{eq:1ptraf}
\end{equation}
We can thus switch to single-particle matrices $h_{n,i} = \exp ( 2 \kappa_{n,i}' Y)$ and $v_{n,i} = \exp (2 \kappa_{n,i} Y)$ instead of the respective many-body operators $H_{n,i}$ and $V_{n,i}$~\cite{Jian:2022jg}, where the  $2\times 2$ Pauli matrix $Y$  acts on the $(2i-1,2i)$\textsuperscript{th} (for $h_{n,i}$) and $(2i,2i+1)$\textsuperscript{th} degrees of freedom (for $v_{n,i}$).
We denote the resulting $2M\times 2M$ transfer matrix by $\Mtot$. 

For real $J$, the single-particle operators are pseudo-unitary, $Z t^{-1} Z = t^\dagger$ with $t= v_{n,i}$ or $t=h_{n,i}$, and can thus readily be interpreted as single-particle transfer matrices~\cite{Merz:2002gj} that, when acting on a pair of counterpropagating modes $\mathbf{c} = (c_n,c_{n+1})$, conserve their current $\mathbf{c}^\dagger Z \mathbf{c}$~\cite{Beenakker:1997gz}, cf.\ Fig.~\ref{fig:network_model}(a).
In this way, each matrix $t= v_{n,i}$ or $t=h_{n,i}$ describes the scattering at a ``junction", and the junctions form a scattering network. 

For complex $J$, neither the $v_{n,i}$ nor the $h_{n,i}$ are pseudo-unitary~\cite{Venn2022}.
However, since $\tilde{\kappa}_{n,i} = i (\phi - (1-\eta_{n,i}^{(h)}) \pi/4)$ is purely imaginary, $h_{n,i}$ is always unitary and we can interpret it as a scattering matrix connecting co-propagating modes.
While $v_{n,i}$ is not unitary, the product $X v_{n,i}$ is pseudounitary:  $Z[ X v_{n,i}]^\dagger Z =[ X v_{n,i}]^{-1}$. That is, when acting on two counterpropagating modes, $v_{n,i}$ swaps them and conserves current~\cite{Venn2022}, cf.\ Fig.~\ref{fig:network_model}(b).

The single-particle transfer matrices for both real and complex $J$ imply real scattering matrices for each junction~\cite{Read:2000hu,Merz:2002gj,Venn2022}. 
This places the networks into Altland and Zirnbauer's symmetry class D~\cite{AZ97}. 
Hence, the links of the networks can be interpreted as describing directed 1D Majorana modes. 
(Upon taking the networks together with their time-reversed partners, the real and complex $J$ cases realize two limits of the  time-reversal symmetric class DIII Majorana network in Ref.~\onlinecite{FulgaDIII}.)

Networks in symmetry class D often include disorder in the form of randomly placed ``vortices"~\cite{Cho:1997bs,Read:2000hu,Chalker:2001he,Merz:2002gj,Evers:2008gi}. 
(A vortex is a point-defect such that a mode encircling it picks up an extra $\pi$ phase.)
In our case the disorder is via the $\eta_{n,i}$ and this indeed introduces vortices. 
In the incoherent case, as is well known from the RBIM~\cite{Read:2000hu,Chalker:2001he,Merz:2002gj}, $\eta_{n,i}=-1$ imprints a pair of vortices adjacent to junction ${n,i}$ in one of the sublattices. 
In terms of the surface code, a vortex appears at the adjacent $S_w^Z$, i.e., where $S_w^Z =-1$ due to the bit flip represented by $\eta_{n,i}=-1$. 
In the coherent case, $\eta_{n,i}=-1$ has the same effect, however, the manner in which a vortex can be encircled is different than in the incoherent case due to the propagation directions being laid out differently in the coherent errors' scattering network. 

The networks, together with the vortex distribution, determine the phase of QEC~\cite{Venn2022}.
Using this fact, we can further substantiate why the $p=\sin^2(
\phi)$ partial Pauli twirl line of our simplified $\phi-p$ model is expected to capture the qualitative phase structure of the full coherent error model:
The network itself is the same for the two models since, apart from the bond signs, they originate from the same complex-$J$ Ising model.
(As we noted in Sec.~\ref{sec:Ising_mappings}, this captures the coherent summation over $X$-strings, the key feature of the coherent error problem.)
By sampling $\eta_{n,i}$ differently, the two models differ in their vortex distribution.
However, for both models, the rarity of $\eta_{n,i}=-1$ for small $\phi$ implies tightly bound vortex pairs, while for sufficiently large $\phi$ vortices proliferate.
These basic features dictate~\cite{Chalker:2001he} that the qualitative phase structure along $p=\sin^2(\phi)$ is the same for the two models, albeit the quantitative details such as the phase boundary $\phi_c$ or the critical properties may differ.

Our characterization of network models will include transport properties, specifically the dimensionless conductivity $g=(L/M) \langle \tr [\mathcal{T}^\dagger \mathcal{T} ] \rangle_\mathrm{dis}$. 
Here $\mathcal{T}$ denotes the transmission matrix from the transmission-reflection grading of the total scattering matrix $\Stot=\left(\begin{smallmatrix}
\Rtot & \Ttot' \\
\Ttot & \Rtot'
\end{smallmatrix}\right)$~\cite{Beenakker:1997gz} and $\langle \dots \rangle_\mathrm{dis}$ denotes the disorder average.
In an insulator, i.e., a localized network, the conductivity satisfies  $g\propto e^{-2L/\xi}$ where $\xi$ is the localization length~\cite{Evers:2008gi}. 
A metallic network, in contrast, displays $g\propto \ln(L)$. 
Both expressions hold in the large $L$ limit, understood to be taken with fixed aspect ratio $L/M$.

\section{Entanglement phases via 2D Ising models, networks, and 1D fermions}

We next discuss how 2D Ising considerations combined with links between 2D scattering networks and 1D free-fermion parent Hamiltonians illuminate the entanglement phases of $\ket{\Psi_\infty}$. 
Our approach in this Section can be generalized to other fermionic quantum circuits, beyond our motivating surface-code problems, and hence may be of independent interest. 
The Ising model and parent Hamiltonian perspectives complement recent tensor-network- and scattering-network-based approaches~\cite{Jian:2022jg} to entanglement phases in free-fermion circuits. 
In Sec.~\ref{sec:results},  we shall numerically confirm the insights we obtain here, returning our focus to the entanglement phases in the quantum circuits dual to the surface code with bit flips and coherent errors.

\subsection{2D networks and 1D parent Hamiltonians}
\label{sec:2Dnet1DH}

We now link some features of 2D networks and of $\mathcal{H}$ from $\Mtotop \Mtotop^\dagger=e^{-L\mathcal{H}}$. 
We follow Ref.~\onlinecite{Venn2022}, where we noted that the links we describe bridge between the approach of Ref.~\onlinecite{Fulga12} relating 1D and 2D topological phases via scattering matrices (the 1D Hamiltonians there, however, arise differently than here) and Ref.~\onlinecite{Merz:2002gj}'s pioneering insights linking topology in 2D networks and 1D systems. 
We focus on purely fermionic $\Mtotop$.

The first key observation is that an insulating (i.e., localized) network implies that $\mathcal{H}=\frac{i}{2}\sum_{ij}a_{ij} \gamma_i\gamma_j$ is gapped. (Here we introduced the single-particle Hamiltonian $ia$ with a real antisymmetric matrix $a$.)
To see this, we note that Eq.~\eqref{eq:1ptraf} implies 
\begin{equation}
\Mtot\Mtot^\dagger=\exp[-2iLa]
\label{eq:MvsA}
\end{equation}
for the matrix $\Mtot$ for $\Mtotop$. 
This links the single-particle energies $\varepsilon_j\ge 0$ of $ia$ to transport properties~\cite{Merz:2002gj}. 
In particular, one can show that the conductivity satisfies
\begin{equation}
g=\frac{L}{M}\left\langle\sum_{j=1}^M\frac{1}{\cosh^2(L{\varepsilon}_j)}\right\rangle_\mathrm{dis}.
\end{equation}
In an insulator, the $g\propto e^{-2L/\xi}$ large-$L$ asymptotics (with fixed $L/M$) implies  $\lim_{M\rightarrow\infty} {\varepsilon}_1>0$ for the smallest energy ${\varepsilon}_1$. 
(The energies $\varepsilon_j$, and as such ${\varepsilon}_1$, become increasingly non-random upon increasing the system size~\cite{Beenakker:1997gz,Merz:2002gj}.)
Hence, $\mathcal{H}$ is gapped, with gap $\lim_{M\rightarrow\infty} {\varepsilon}_1 = \alpha \xi^{-1}$ (with $\alpha > 0$ order of unity accounting for the difference between average and typical $\xi$~\cite{Evers:2008gi}).
In what follows, we refer to gapped $\mathcal{H}$ and insulating networks interchangeably.

The second key link, also implied by Eq.~\eqref{eq:MvsA}, is between the 1D topological invariant $\mathcal{I}$ and the reflection matrix $\Rtot'$ of the 2D scattering network. Specifically, one can show that for a gapped $\mathcal{H}$, i.e., an insulating network, we have $\mathcal{I}=\sgn[\det(\Rtot'_\text{pbc}\Rtot'_\text{apbc})]$~\cite{Venn2022}. 
(Replacing $\Rtot'$ by $\Rtot$ gives the same result~\cite{Fulga12}.)

\subsection{Area law phases}
\label{sec:ALth}

We now show that when the purely fermionic $\mathcal{H}$ is gapped, i.e., the corresponding network is insulating, then $\ket{\Psi_\infty}$ satisfies the entanglement area law. 
If we knew that $\mathcal{H}$ is a local Hamiltonian, this would be an immediate consequence of its gap~\cite{Hastings:2007bu,Eisert:2010hq}. 
However, from $\Mtotop \Mtotop^\dagger=e^{-L\mathcal{H}}$ the locality is not obvious, even if $L\propto$ the circuit depth of $\Mtotop$ makes it plausible. 
To establish the area law, we will show that the correlations $\bra{\Psi_\infty}\gamma_a \gamma_b \ket{\Psi_\infty}$ decay exponentially with $|a-b|$ (for $\xi\ll|a-b|\ll M$); this provides a sufficient condition for $\ket{\Psi_\infty}$ to display an area law~\cite{Brandao:2013fz,Brandao:2015cn,Cho_PhysRevX.8.031009}. 
Our approach does not assume the absence of disorder from $\eta_{n,i}$; in this way it  complements the analytical arguments in Ref.~\cite{Jian:2022jg} based on disorder-free networks. 

We start by noting that for large $L$
\begin{equation}
-i C_{ab}^{(\infty)} = \bra{\Psi_\infty}\gamma_a \gamma_b \ket{\Psi_\infty}\to \frac{\text{Tr} \Mtotop^\dagger \gamma_a \gamma_b \Mtotop}{\text{Tr} \Mtotop^\dagger \Mtotop}, 
\end{equation}
where we take the trace in terms of the bosonic problem (i.e., use $P$-dependent boundary conditions)~\cite{Trfn}. 
This allows us to view $\bra{\Psi_\infty}\gamma_a \gamma_b \ket{\Psi_\infty}$ as an Ising correlator on the torus. 
This enables the use of space-time duality~\cite{Ippoliti:2021fu,Lu:2021da} to evaluate the correlation function.

The corresponding Ising model is defined by $\Mtotop^\dagger \Mtotop$; it consists of two coupled Ising patches, one for $\Mtotop^\dagger$ and one for  $\Mtotop$. 
These two Ising patches are in the same phase: the Hamiltonian for $\Mtotop$ and $\Mtotop^\dagger$ have identical spectra, so both of them are gapped; the corresponding phases are labeled by $\mathcal{I}$ for $\ket{\Psi_\infty}$. By $\mathcal{I}$ being defined by the fermion parity, and the parity being the same for $\ket{\varphi_\text{min}}$ and $\ket{\widetilde{\varphi}_\text{min}}$, the value of $\mathcal{I}$ for $\Mtot$ is the same as for $\Mtot^\dagger$. 

The correlation function is thus that of $\gamma_a \gamma_b$ embedded in the bulk in the transfer matrix of this 2D Ising model. 
To interpret this in the Ising language, we take $a<b$ without loss of generality, and implement $\gamma_a\gamma_b=\gamma_a\gamma_{a+1}\gamma_{a+1}\gamma_{a+2}\ldots\gamma_{b-1}\gamma_b$, up to an overall phase, by $\kappa\to\kappa+i\pi/2$, $\tilde{\kappa}\to\tilde{\kappa}+i\pi/2$ in the last layers of $\Mtotop$, while leaving $\Mtotop^\dagger$ unchanged.
This introduces $J \eta_{n,i}^{(v)}\to J \eta_{n,i}^{(v)}+i\pi/2$ and $J \eta_{n,i}^{(h)}\to -J \eta_{n,i}^{(h)}$ along the line from $a$ to $b$. 
In the 2D Ising language, the former yields $\sigma_a \sigma_{b}$, while the latter yields a seam of flipped horizontal bonds from $a$ to $b$. 
The corresponding correlator is that of products of Ising spins and disorder operators: an Ising fermion correlator~\cite{fradkin2017disorder}. 
This decays exponentially for both $\mathcal{I}=\pm1$ due to either the disorder or the Ising correlators decaying exponentially while the other being constant~\cite{Merz:2002gj,Venn2022,fradkin2017disorder}.
Using space-time duality to orient the fermion string for $\gamma_a\gamma_b$ along the temporal direction, one can show that $C_{ab}^{(\infty)} \propto e^{-\alpha |a-b|/2\xi}$, with $\xi$ the localization length in the scattering network.
This holds both typically and on average because the Ising model, or network, for $\Mtotop^\dagger\Mtotop$ has $\eta=-1$ strings appear in pairs thus the rare long $\eta=-1$ strings (cf.\ Sec.~\ref{sec:GGSBC}) in the dual-temporal direction are inoperative.

This establishes the $\mathcal{I}=\pm1$ gapped phases of $\Mtotop$, and the respective insulating phases of the scattering networks, as yielding an area-law $\ket{\Psi_\infty}$. 
The exponentially decaying correlations also imply that $iC^{(\infty)}$ is a quasilocal (i.e., with couplings exponentially decaying with distance) single-particle Hamiltonian; it has eigenvalues $\pm1$ and hence defines a gapped quasilocal parent Hamiltonian for $\ket{\Psi_\infty}$~\cite{Fidkowski:2010ck,Beri11,Yin19}.
This results in the following signatures for the single-particle entanglement spectrum~\cite{Fidkowski:2010ck,Turner:2011gp}: 
For both $\mathcal{I}=\pm1$, the entanglement Hamiltonian $i\bar{C}_{M/2}^{(\infty)}$ has a bulk ``entanglement gap". 
When $\mathcal{I}=1$, the entire single-particle entanglement spectrum is gapped.
When $\mathcal{I}=-1$, however, the nontrivial topology implies entanglement zero modes (analogous to Majorana end states at physical boundaries). 
For finite $M$, the zero modes are split, yielding an entanglement energy level $\lambda_0$ satisfying $\lambda_{0}\propto e^{-M/c}$ with $c>0$ increasing with $\xi$.

\subsection{Logarithmic entanglement phases}
\label{sec:logEE}

A gapless $\mathcal{H}$ can also arise; this happens if the network is metallic. 
While this is ruled out for an Ising model with real couplings~\cite{Read:2000hu}, a metallic phase is generically part of the phase diagram when the couplings are complex~\cite{Chalker:2001he}. 
In this case, from $\ket{\Psi_\infty}$ being the ground state of a gapless 1D $\mathcal{H}$, by analogy to the logarithmic $S_{M/2}$ at criticality~\cite{Vidal03,calabrese2004entanglement,Refael04,Li:2018bf,Skinner:2019fn,LiPRBet19,Jian:2020cp,LiCFT21} we expect $S_{M/2}\propto \ln M$, i.e., a logarithmic entanglement phase. (See also Ref.~\onlinecite{Jian:2022jg} for linking metallic networks to logarithmic entanglement phases.)

A prediction on scaling beyond these asymptotics can also be made if we note that the physics of metallic 2D networks is described by a nonlinear $\sigma$ model~\cite{Read:2000hu,Evers:2008gi}. 
(We numerically verify this in Sec.~\ref{sec:results} for the coherent-error network.)
In this model, the conductivity $g$ is the only coupling; as a consequence, it follows single-parameter scaling $g(L;p,J)=g[L/\ell_{p,J}]$  tracing out the renormalization group flow of $g$. 
(Here $\ell_{p,J}$ is an effective length scale.)
This is a characteristic feature of the metallic phase that holds beyond the asymptotic  $g\propto \ln L$ regime. 
(The key requirement is diffusive transport, setting in for $L$ much larger than the mean free path, i.e., the short-distance cutoff for the nonlinear $\sigma$ model.)
Based on this, we similarly expect single-parameter scaling for the entanglement, $S_{M/2}(p,J)=S(M/m_{p,J})$, providing an entanglement fingerprint of the nonlinear $\sigma$ model. 

\section{Numerical results}
\label{sec:results}

We now return to the link between the phases of QEC in the surface code and the entanglement phases of their dual quantum circuits. 

\subsection{Real Ising couplings}

For the real-coupling RBIM, and thus the surface code with random bit flips, the transport properties of the network model have been extensively discussed in the literature~\cite{Chalker:2001he,Motrunich:2001gl,Merz:2002gj,Merz:2002ia,Evers:2008gi}.
Here, we highlight one key observation: The phases on both sides of the transition are insulating, but characterized by different topological invariants~\cite{Merz:2002gj,Venn2022}: we have $\mathcal{I}= -1$ in the ordered Ising phase (including the error-correcting part of the Nishimori line) and $\mathcal{I}=1$ otherwise, as shown Fig.~\ref{fig:phase_diagram}(a).

\begin{figure}
\includegraphics[scale=1]{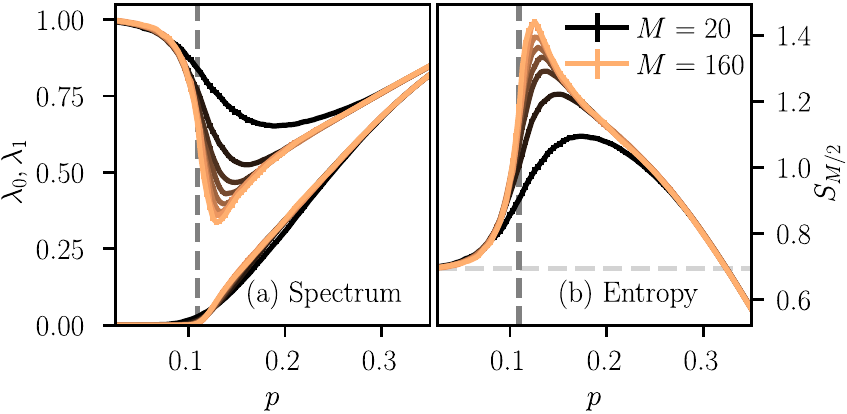}
\caption{(a) Entanglement spectrum and (b) entanglement entropy for real $J$ on the Nishimori line. 
We use $L=5M$ and cylinder circumferences from $M=20$ (black) to $M=160$ (light orange). We averaged over 16--128 configurations of $\eta$; error bars (2$\times$standard error) are imperceptible.
The $p=0.1093$  vertical dashed line marks the Nishimori point~\cite{Merz:2002gj}, i.e.,  the QEC threshold~\cite{Dennis:2002ds}. The horizontal dashed line in panel~(b) marks the $\log 2$ bound in the $\mathcal{I}=-1$ phase. }
\label{fig:entanglement_inc}
\end{figure}

Turning to entanglement, in Fig.~\ref{fig:entanglement_inc} we show the entanglement spectrum and entropy $S_{M/2}$ along the Nishimori line (i.e., for surface-code QEC).
Our initial state $\ket{\Psi_0}$ is a random half-filled state [defined in terms of fermions $c_j = (\gamma_{2j-1}+ i \gamma_{2j})/2$], which we evolve for long cylinders, $L=5M$.
We find that the entanglement spectrum and entropy converge, indicating that $\ket{\Psi_\infty}$ has been reached.

In the entanglement spectrum, we observe the following features: Below the error threshold, $p_c \approx 0.1093$~\cite{Merz:2002gj,Dennis:2002ds}, where $\mathcal{I}=-1$, the single-particle entanglement spectrum is gapped and has a zero mode whose energy decays exponentially with system size; the many-body entanglement spectrum is thus degenerate in the infinite-system limit.
These features confirm expectations from Sec.~\ref{sec:ALth} for a topologically nontrivial phase. 

The smallest entanglement eigenvalue $\lambda_1$ of the bulk is minimal close to the transition.
With increasing $M$, the $p$ where $\lambda_1$ is minimal shifts towards $p_c$ and the minimum itself $\left. \lambda_1 \right|_\mathrm{min}$ decreases as a power law with $M$, consistent with a critical phase at the transition~\cite{Merz:2002ia}.

On both sides of the transition, the entanglement entropy scales as an area law, i.e., it does not increase with the system width $M$.
This again confirms expectations from Sec.~\ref{sec:ALth} for $\ket{\Psi_\infty}$ associated to insulating networks. 

For $\mathcal{I}=-1$, the entanglement entropy is bound from below by $\log 2$, which reflects the presence of a zero mode. 
For $\mathcal{I}=1$, the entropy $S_{M/2}$ goes to zero for large $M$ and sufficiently large $p$.
Near the transition, the entanglement entropy grows with $M$.
Consistently with the area law away from $p_c$, this is expected saturate unless $p=p_c$. 
This is consistent with the $p$ for which $S_{M/2}$ is maximal shifting towards $p_c$ with increasing $M$.

\subsection{Complex Ising couplings}

\begin{figure}
\includegraphics[scale=1]{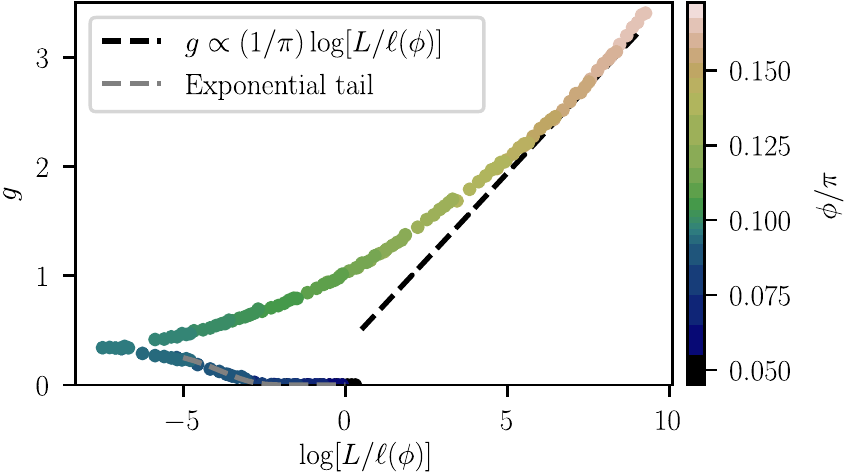}
\caption{Dimensionless conductivity $g$ for complex couplings at $p=\sin^2 \phi$ as a function of the rescaled system length $L/\ell(\phi)$ for wide systems with $M=5L$, averaged over 100--$10^{4}$ configurations of $\eta$; error bars (2$\times$standard error) are imperceptible.
For angles above a critical $\phi_c$, the conductivity increases with $L/\ell(\phi)$ (metallic phase; dashed black line shows $g \propto (1/\pi) \log [L/\ell (\phi)]$). Below the transition, it decreases exponentially to zero (localized phase, dashed gray line shows the exponential tail).}
\label{fig:conductivity_twirl}
\end{figure}

In Fig.~\ref{fig:conductivity_twirl}, we show the  conductivity $g$ for the complex RBIM motivated by coherent errors, focusing on the $p=\sin^2 \phi$ partial Pauli twirl  line [shown dashed in Fig.~\ref{fig:phase_diagram}(b)].
To probe the bulk value of $g$ we work with a wide cylinder, $M=5L$~\cite{Medvedyeva:2010jf,FulgaDIII}. 
The results for different $\phi$ are shown with different colors.
When rescaling the length to a dimensionless $L/\ell(\phi)$ with an appropriately chosen function $\ell (\phi)$, the conductivity data collapses onto one of two scaling curves, depending on $\phi$. (For completeness, we show the unscaled data in Appendix~\ref{sec:appendix_raw}.)

For angles $\phi > \phi_c$, the system is metallic: $g$ increases with $L$ and for sufficiently large systems it approaches the universal class-D result~\cite{Evers:2008gi} $g \propto (1/\pi) \ln L$ (dashed black line).
For $\phi<\phi_c$ the system is in an insulating phase: for large systems, $g$ decreases exponentially with $L$ (dashed gray line). In this phase, $\ell (\phi)$ is the localization length; it diverges close to the transition. 
The metal-insulator transition occurs at $\phi_c = (0.095 \pm 0.005)\pi$---note that this value is significantly smaller than the coherent error threshold $\phi_\mathrm{th} = (0.14\pm 0.005)\pi$ we found in Ref.~\onlinecite{Venn2022} by sampling the syndromes according to $P_s$ instead of sampling each $\eta$ independently as we do here.

In the insulating phase, we find $\mathcal{I}=-1$, as in the ordered Ising insulator for real $J$. (Our results are also consistent with the $\mathcal{I}=-1$ insulator for vortices sampled according to $P_s$~\cite{Venn2022}.)
On leaving the asymptotics, the scaling curves we find are qualitatively similar to previous results for class-D metal-insulator transitions~\cite{Medvedyeva:2010jf,Wang:2021jn,Venn2022}. 
Furthermore, the scaling in the metallic regime follows closely the nonlinear $\sigma$ model renormalization group flow for $g$~\cite{Evers:2008gi}. 
This excellent agreement with nonlinear $\sigma$ model predictions is in contrast to the results for $\eta$ being sampled according to $P_s$ for coherent errors~\cite{Venn2022}.

\begin{figure}
\includegraphics[scale=1]{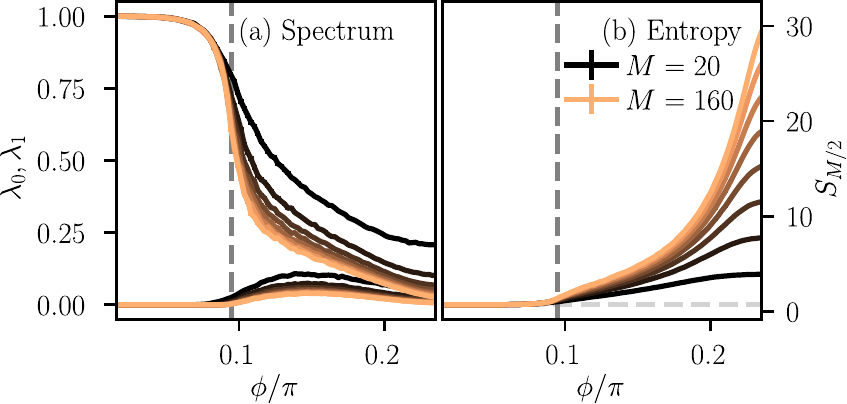}
\caption{(a) Entanglement spectrum and (b) entanglement entropy for complex $J=-(1/2) \log (i \tan \phi)$ and $p = \sin^2 \phi$ for $L=5M$ and cylinder circumferences from $M=20$ (black) to $M=160$ (light orange). We averaged over $2^5$--$2^8$ configurations of $\eta$; error bars (2$\times$standard error) are imperceptible.
The $\phi=0.095\pi$ dashed line marks the entanglement transition; the horizontal dashed line in panel~(b) marks $S = \log 2$.}
\label{fig:entanglement_coh}
\end{figure}

We now discuss the signatures of these phases in the entanglement spectrum and entropy.
In Fig.~\ref{fig:entanglement_coh}, we show these quantities, continuing to focus on the partial Pauli twirl line $p=\sin^2 \phi$.
We again start the evolution from a random half filled state and converge to $\ket{\Psi_\infty}$ using  long cylinders with $L=5M$.

\begin{figure}
\includegraphics[scale=1]{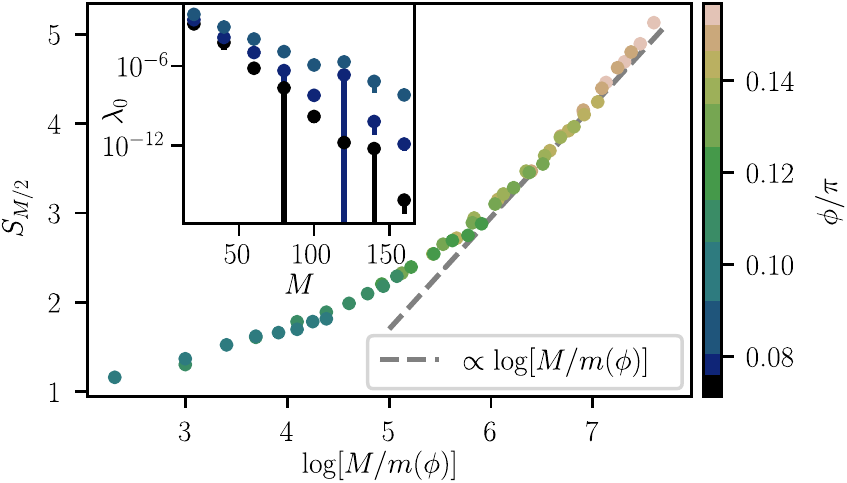}
\caption{Entanglement entropy $S_{M/2}$ in the metallic phase ($\phi>\phi_c$) for complex couplings at $p=\sin^2 \phi$ as a function of the rescaled circumference $M/m(\phi)$. 
We averaged over $2^5$--$2^8$ configurations of $\eta$; error bars are 2$\times$standard error and the gray dashed line serves as a guide for the eye showing a logarithmic increase.
The inset shows $\lambda_0$, the exponentially decaying zero mode; the decay length increases with $\phi$.}
\label{fig:entanglement_twirl_rescaled}
\end{figure}

In the insulating phase, the entanglement spectrum  displays a zero mode and has a bulk gap.
The entanglement entropy displays an area law and it slowly decreases with $\phi$ to the $\phi=0$ value $S_{M/2} = \log 2$.
The entanglement zero mode $\lambda_0$ decays exponentially with $M$, shown in the inset of Fig.~\ref{fig:entanglement_twirl_rescaled} for various angles $\phi<\phi_c$. 
These features agree with the behavior expected for an $\mathcal{I}=-1$ insulator, i.e., a topological area-law $\ket{\Psi_\infty}$, cf.\ Sec.~\ref{sec:ALth}. 

In the metallic phase, the entanglement spectrum gap decreases as a power law  in $M$ (with a $\phi$-dependent power) and the entanglement entropy increases with $M$.
The large-$M$ asymptotic is $S_{M/2}\propto\ln M$ (shown dashed in Fig.~\ref{fig:entanglement_twirl_rescaled}), indicating a logarithmic entanglement phase.
(The data fit $S_{M/2}\propto\ln^2 M$, derived in a related context~\cite{Fava:2023jo}, similarly well.)
Similarly to $g$, rescaling $M \to M/m(\phi)$ by a $\phi$-dependent length $m(\phi)$ collapses data points a smooth curve, shown in Fig.~\ref{fig:entanglement_twirl_rescaled}. 
This confirms the expectations from Sec.~\ref{sec:logEE}: $S_{M/2}$ in the logarithmic entanglement phase inherits single-parameter scaling from $g$. 
[The function $m(\phi)$, however, does not equal $\ell (\phi)$ used for $g$ in Fig.~\ref{fig:conductivity_twirl}.]

\section{Conclusion}

In this work, we related the phases of surface-code QEC for coherent and incoherent errors to entanglement phases.
In particular, using a mapping to a RBIM with real couplings for incoherent~\cite{Dennis:2002ds} and complex couplings for coherent errors~\cite{Venn2022}, we could interpret the RBIM transfer matrix as a quantum circuit for mixed real-imaginary time Gaussian evolution that converges to a long-time Gaussian state $\ket{\Psi_\infty}$ from generic  (e.g., random) definite-parity initial states.

For both error types, the QEC phase is dual to phase where $\ket{\Psi_\infty}$ satisfies the entanglement area law. 
This phase is topologically nontrivial ($\mathcal{I}=-1$), which implies that its gapped single-particle entanglement spectrum supports a zero mode. 
Consequently, the entanglement entropy is bounded from below by $\log 2$. 
Above threshold, and for incoherent errors, we find an $\mathcal{I}=1$ area law. 
The state $\ket{\Psi_\infty}$ again has gapped entanglement spectrum but without a zero mode, and the entanglement entropy approaches zero away from the transition between the two area-law phases.
For coherent errors, we find a logarithmic entanglement phase above the threshold.

The duality between QEC codes and entanglement phases provides a new perspective from which to study the dynamics of hybrid quantum circuits that is entirely distinct from previously considered emergent QEC in hybrid circuits~\cite{hayden2007black,Choi:2020bs,Gullans:2020eg}.
In particular, the surface code with coherent errors provides a natural physical system in terms of which to interpret hybrid dynamics alternating real-time and imaginary-time evolution and the associated transitions between area-law and logarithmic entanglement phases. 
In this sense, it is tempting to think of the logical error rate---a direct indicator of which phase of QEC the system is in---as an indirect fingerprint of entanglement phases and transitions, albeit via the dual system: the QEC code. 

Our results not only show that such hybrid circuits can be motivated by QEC, but the entanglement phases also offer a novel characterization for the phases of QEC.
The area law for the QEC phase is especially important in this regard~\cite{Napp:2022cf}. 
While we demonstrated this area law only for the specific error models we studied, previous results on more general incoherent errors suggest~\cite{Bravyi:2014ja} that the entanglement entropy continues to exhibit an area law in the QEC phase for a broader class of errors. 
This suggests that the quantum circuits dual to the QEC problem---which has more complicated statistical mechanics models for more general errors~\cite{Chubb:2021cn}---can be efficiently simulated in the QEC phase using matrix product states~\cite{Hauschild:2018bp,Cirac:2021gx}.
Using this, and generalizing our approach to deriving statistical mechanics models for coherent errors, one may chart out the QEC phase for a broad class of errors, including the important open problem of coherent errors with generic SU($2$) rotations~\cite{Kueng16,Wallman16,Debroy18,Bravyi:2018ea,iverson2020coherence,Hashim21}.

Our analysis using scattering networks also offers new perspectives on the relations between such network models and entanglement~\cite{Jian:2022jg}.
Our Ising considerations link insulating networks and  area-law phases explicitly via correlations; this complements existing arguments~\cite{Jian:2022jg} based on disorder-free networks. 
The link to quasilocal parent Hamiltonians and their topological invariants are, to our knowledge, also new aspects connecting scattering networks and entanglement phases~\cite{Jian:2022jg}. 
The entanglement gap and the presence or absence of entanglement zero modes emerge directly and naturally in this approach. 
The link between metals and logarithmic entanglement phases we find agrees with Ref.~\onlinecite{Jian:2022jg}. 
To elucidate this link further, we showed that the entanglement entropy follows single-parameter scaling, similarly to the conductivity. 
The very good agreement we found for the latter with nonlinear $\sigma$ model predictions  suggests that a $\sigma$-model theory may be developed also for the entanglement entropy.
(See Refs.~\onlinecite{Fava:2023jo,Poboiko:2023bm}, that appeared independently of this work, for such $\sigma$-model theories.)

Our results can also be viewed as pertaining to error-corrupted topological quantum memories. Unlike Refs.~\onlinecite{Bao2023,Fan2023,Lu:2023ep,Zou:2023he}, that appeared independently of this work, we focus on the state post stabilizer measurement, and encode all stabilizer information in a $(1+1)$D circuit.
The circuit can be interpreted as the boundary theory of the error-corrupted state and the phase transition of this boundary theory expresses the loss of topologically encoded information in the bulk~\cite{Bao2023,Fan2023}.

The quantum circuit duals for the surface code problems we study display some analogies to a family of Gaussian fermionic circuits studied recently~\cite{Nahum:2020kj,Lang:2020kg,Sang:2021fl,Sang:2021dg,Turkeshi:2021ik,Turkeshi:2022dv,Merritt2023}. 
It would be interesting to generalize our approach to construct network and Ising models for these circuits and thereby to characterize the ``gapped" (area law) and ``Goldstone" (logarithmic entanglement) phases found in their hybrid~\cite{Sang:2021dg,Turkeshi:2021ik,Turkeshi:2022dv,Merritt2023} and measurement-only~\cite{Nahum:2020kj,Lang:2020kg,Sang:2021fl} variants.
Network models may shine light on the classification of these area-law phases, including explicitly establishing the topological origin of the $\log 2$ entanglement entropy arising in one of these phases~\cite{Nahum:2020kj,Lang:2020kg,Merritt2023}.

\begin{acknowledgments}
This work was supported by EPSRC grant EP/V062654/1, a Leverhulme Early Career Fellowship, the Newton Trust of the University of Cambridge, and in part by the \mbox{ERC Starting Grant No. 678795 TopInSy}.

Our simulations used resources at the Cambridge Service for Data Driven Discovery operated by the University of Cambridge Research Computing Service (\href{https://www.csd3.cam.ac.uk}{www.csd3.cam.ac.uk}), provided by Dell EMC and Intel using EPSRC Tier-2 funding via grant EP/T022159/1, and STFC DiRAC funding (\href{https://www.dirac.ac.uk}{www.dirac.ac.uk}).
\end{acknowledgments}

\appendix

\section{Evolution of correlation matrix}
\label{sec:appendix_evolution}

In this appendix, we describe the evolution of an initial Gaussian state via $H_n$ and $V_n$ using the methods outlined in Ref.~\onlinecite{Bravyi:2005jh}.
In particular, instead of the evolution of the density matrix
\begin{equation}
 \rho^{(n+1)} = \frac{T_n \rho^{(n)} T_n^\dagger}{\tr [ T_n \rho^{(n)} T_n^\dagger ]} ,
\end{equation}
where $T_{2n} = H_n$ and $T_{2n-1} = V_n$, we consider the evolution of the correlation matrix $C^{(n)}$ governed by~\cite{Bravyi:2005jh}
\begin{equation}
 C^{(n+1)} = B^{(n)} (1 - C^{(n)} A^{(n)})^{-1} C^{(n)} (B^{(n)})^T + A^{(n)} .
 \label{eq:correlation_evolution}
\end{equation}
The matrices $A^{(n)}$ and $B^{(n)}$ follow from the transformation of the density matrix $\rho_I = 2^{-2M} \prod_a (1 + i \gamma_a \gamma_{a+2M} )$---cf.\ Ref.~\onlinecite{Bravyi:2005jh} for more details.
For a non-unitary evolution
\begin{equation}
 T = \exp ( i z \gamma_j \gamma_k)
\end{equation}
with complex $z$, $A_{jk} (T)=-A_{kj} (T)= \tanh (2 \Re z)$ (and zero for all other entries) and $B (T)$ equals the identity apart from the $2 \times 2$ sector spanned by the $j$\textsuperscript{th} and $k$\textsuperscript{th} indices with $B_{jj} (T) = B_{kk} (T) = \cos (2 \Im z)/\cosh (2 \Re z)$ and $B_{jk} (T)  = - B_{kj} (T)= i \sin (2 \Im z)/\cosh (2 \Re z)$.
Thus, the corresponding matrices for $H_n$ and $V_n$ are block-diagonal with $2\times 2$ blocks
\begin{align}
 A_{n,i} = i \tanh (2\Re z Y) , & &
 B_{n,i} = \frac{\exp \left(- 2 i \Im z Y \right)}{\cosh ( 2\Re z )} .
\end{align}
that for $H_{n,i}$ act on the $(2i-1,2i)$\textsuperscript{th} degrees of freedom with $z=\tilde{\kappa}_{n,i}$ and for $V_{n,i}$ act on the $(2i,2i+1)$\textsuperscript{th} degrees of freedom with $z=\kappa_{n,i}$.

\begin{figure}
\includegraphics[scale=1]{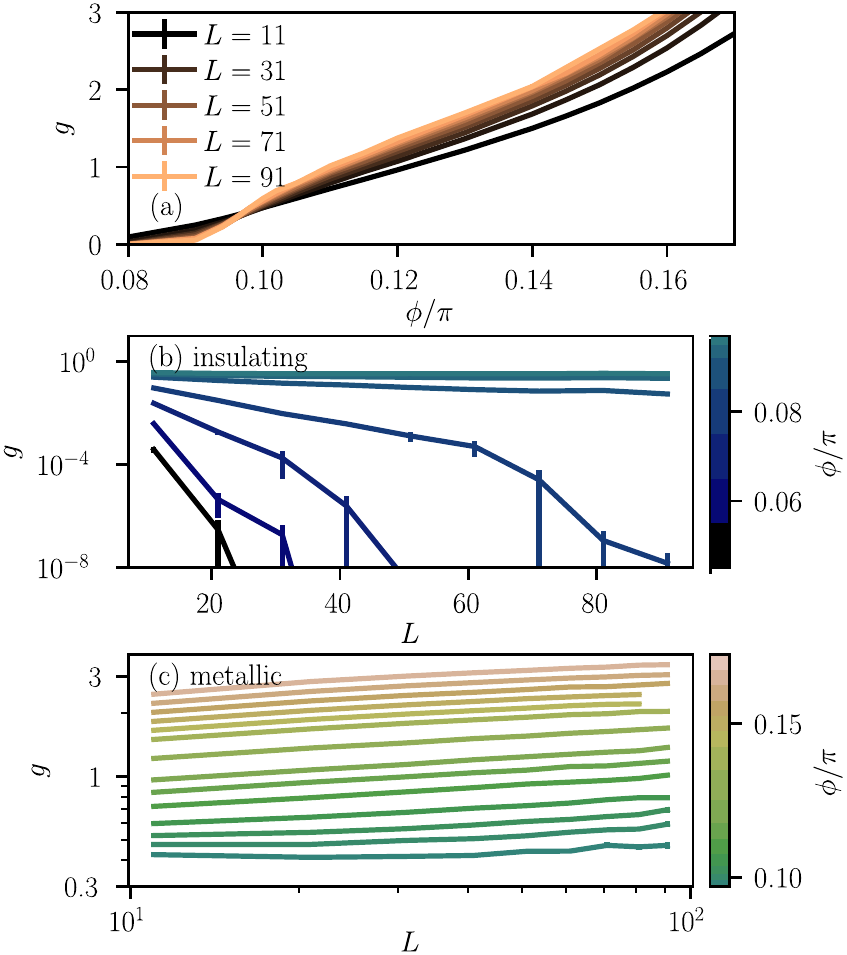}
\caption{Dimensionless conductivity $g$, averaged over 100--$10^{4}$ configurations of $\eta$ (error bars are 2$\times$standard error), as a function (a) of $\phi$ for different sizes and (b)--(c) of $L$ for different angles $\phi$. In (b), we show the conductivity for the insulating case, and in (c) the metallic case.}
\label{fig:conductivity_twirl_raw}
\end{figure}

Instead of considering the consecutive evolution of the correlation matrix via Eq.~\eqref{eq:correlation_evolution}, we can instead fully evolve the state by the purely fermionic transfer matrix $\Mtotop$ [Eq.~\eqref{eq:Mtotop}], which reduces to its single-particle form $\mathcal{M}$ when considering individual Majoranas, cf.\ Eq.~\eqref{eq:1ptraf}.
We first consider real couplings $J$.
The polar decomposition~\footnote{It is numerically more stable to transform the product of transfer matrices into a composition of scattering matrices~\cite{Tamura:1991ki}.} of the product of transfer matrices $\mathcal{M}' = Q \mathcal{M} Q^\dagger$ with $Q = (1/\sqrt{2}) ( 1 + i \sigma_z)$ is~\cite{Mello:1988cj,Beenakker:1997gz}
\begin{equation}
 \mathcal{M}' = \begin{pmatrix} v & \\ & {v’}^T \end{pmatrix} \begin{pmatrix} \cosh \mathcal{D} & \sinh \mathcal{D} \\ \sinh \mathcal{D} & \cosh \mathcal{D} \end{pmatrix} \begin{pmatrix} u’ & \\ & {u}^T \end{pmatrix} 
\end{equation}
where $v,v',u,u'$ are orthogonal matrices since the network is in symmetry class D, and $\mathcal{D} = \diag ( \{ L  \varepsilon_j'\})$ with $|\varepsilon_j'|=\varepsilon_j$, cf.\ Eq.~\eqref{eq:MvsA}.
The determinant of the reflection matrix $\mathcal{R} = - u \tanh \mathcal{D} u'$ is $\det \mathcal{R} = (-\tanh (L \varepsilon_j'))^M \det u \det u'$.

We choose $\varepsilon_{j>0}'>0$ and $\mathrm{sgn}(\varepsilon_0') = (-1)^M \det \mathcal{R}$ to ensure $\det u \det u' = 1$. This automatically fixes $\det v \det v' = 1$.
Since their determinants equal $1$, the block-diagonal matrices $\diag (u', u^T) = \exp (h_u)$ and $\diag (v,{v'}^T) = \exp (h_v)$~\cite{Kitaev:2001gb} with real antisymmetric $h_{v/u}^T = - h_{v/u}$.
The transfer matrix $\mathcal{M} = Q^\dagger \mathcal{M}' Q$ is accordingly a product of exponentials
\begin{equation}
 \mathcal{M} = \underbrace{\begin{pmatrix} v & \\ & {v'}^T \end{pmatrix}}_{\exp (h_v)} \underbrace{\begin{pmatrix} \cosh \mathcal{D} &  -i \sinh \mathcal{D} \\ i \sinh \mathcal{D} & \cosh \mathcal{D} \end{pmatrix}}_{\exp (\mathcal{D} Y)} \underbrace{\begin{pmatrix} u' & \\ & {u}^T \end{pmatrix}}_{\exp (h_u)} .
\end{equation}
The corresponding many-body operators $U = \exp ( \sum_{jj'} [h_u]_{jj'} \gamma_{j} \gamma_{j'}/4 )$, $V = \exp ( \sum_{jj'}  [h_v]_{jj'} \gamma_{j} \gamma_{j'} /4)$, and $D = \exp ( -i L \sum_j \varepsilon_j' \gamma_j \gamma_{j+M}/2)$ can be straightforwardly implemented in fermionic linear optics~\cite{Bravyi:2005jh}.
Thus, the evolution of the correlation matrix $C^{(2L)}$ requires only three steps.

For complex Ising couplings, only multiples of four layers are current-conserving and thus be decomposed as scattering matrices [cf.\ Fig.~\ref{fig:network_model}(b) in the main text]. For these current-conserving sequences (even $L$), the same steps described above can be used; when $L$ is odd, we additionally need to evolve the correlation matrix by the remaining $H_n$ and $V_n$.

\section{Raw conductivity data}
\label{sec:appendix_raw}

In the main text, we show the dimensionless conductivity $g$ as a function of the rescaled system size $L/\ell (\phi)$ [Fig.~\ref{fig:conductivity_twirl}].
For completeness, we show the raw data without the $\phi$-dependent rescaling $\ell (\phi)$ in Fig.~\ref{fig:conductivity_twirl_raw}.
In panel~(a), $g$ as a function of $\phi$ for different system sizes. For small angles $\phi<\phi_c$ in the insulating regime, the conductivity decreases with $L$, and for angles above the transition, $g$ increases with $L$.
In panels~(b) and~(c), we show $g$ as a function of $L$ for various angles, where we split up the data into the insulating regime [panel~(b)] and metallic regime [panel~(c)]. Note panel (b) uses a log-scale (for better visibility of the exponential decay) and (c) a log-log scale (for better visibility of power laws at large $\phi$).

\bibliography{references}

\begin{thebibliography}{125}
\bibitem [{Trf()}]{Trfn}%
  \BibitemOpen
  \href@noop {} {}\bibinfo {note} {When $\mathcal{I}=-1$, the trace gets
  contributions from both $\ket{\Psi_\infty}$ and its opposite parity partner
  ground state; the result gives $\bra{\Psi_\infty}\gamma_a \gamma_b
  \ket{\Psi_\infty}$ because these are equal contributions up to corrections
  exponentially small in $M$}\BibitemShut {NoStop}%
\end{thebibliography}

\end{document}